\documentclass[onecolumn,prd,amsmath,amsssymb,groupedaddress,12pt]{revtex4}
\usepackage{graphicx}
\usepackage{dcolumn}
\usepackage{bm}
\usepackage{amssymb}
\usepackage{dsfont}
\def\ba{\begin{eqnarray}}
\def\ea{\end{eqnarray}}
\def\be{\begin{equation}}
\def\ee{\end{equation}}

\def\nn{\nonumber}

\def\d{\mathrm{d}}

\def\mn{_{\mu \nu}}

\def\({\left(}
\def\){\right)}

\def\nn{\nonumber}
\def\ms{M_6^4{}}
\def\mf{M_5^3{}}
\def\mq{M_4^2{}}

\def\bN{\bar N}
\def\Ny{\bar{N}_y}
\def\Ln{\mathcal{L}_n}

\def\py{\partial_y}
\def\ts{\tilde \sigma}

\begin{document}

\date{\today}

\title{Cascading Gravity is Ghost Free}

\author{Claudia de Rham$^{1}$,  Justin Khoury$^{2}$, Andrew J. Tolley$^{3}$}

\affiliation{
\nobreak{$^1$D\'epartement de Physique Th\'eorique, Universit\'e de Gen\`eve,
24 Quai E. Ansermet, CH-1211 Gen\`eve,
Switzerland}\\
\nobreak{$^2$Center for Particle Cosmology, University of Pennsylvania, Philadelphia, PA 19104-6395, USA}\\
\nobreak{$^3$Perimeter Institute for Theoretical Physics, 31 Caroline St. N., Waterloo, ON, N2L 2Y5, Canada}}

\begin{abstract}
\begin{center}
{\bf Abstract}
\end{center}
\noindent
We perform a full perturbative stability analysis of the 6D cascading gravity model in the presence of 3-brane tension. We demonstrate that for sufficiently large tension on the (flat) 3-brane, there are no ghosts at the perturbative level, consistent with results that had previously only been obtained in a specific 5D decoupling limit. These results establish the cascading gravity framework as a consistent infrared modification of gravity.
\end{abstract}
\maketitle

\section{Introduction}

Ten years after its discovery, cosmic acceleration remains one of the central puzzles in cosmology.
Although the data is thus far consistent with dark energy plus General Relativity (GR), a tantalizing alternative is that new
gravitational degrees of freedom on cosmological scales are responsible for late-time acceleration.
The most compelling idea along these lines is that the graviton has a small mass or width, of the order of today's
Hubble parameter, which could account for the apparent smallness of the cosmological constant via the degravitation mechanism~\cite{dgs,addg,degrav}.

The central difficulty in constructing any consistent infrared (IR) modified theory of gravity is the avoidance of ghosts. For instance, massive gravity theories
with a hard mass for the graviton~\cite{FP} are well-known to be unstable~\cite{BoulwareDeser,nima,Creminellipaper}. More promising constructions
rely on branes and extra dimensions~\cite{DGP,cascade1,cascade2,deRham:2008qx,intersecting,cascade3,aux1,aux2}, such as the
Dvali-Gabadadze-Porrati (DGP) model~\cite{DGP}. The normal branch of the DGP model is perturbatively ghost free, in contrast to
the self-accelerating branch~\cite{lpr,dgpghost}, and thus represents a perturbatively consistent infrared modification of gravity in which the gravity has
a soft mass, {\it i.e.}, it is a resonance.

The Cascading Gravity framework~\cite{cascade1,cascade2,deRham:2008qx,intersecting,cascade3} proposed recently generalizes the DGP model to
higher dimensions. There are two motivations for considering this broader set-up.
As theoretical motivation, it offers a new class of IR modified gravity theories whose properties are therefore worth exploring. In particular, the
lower momenta dependance of the mass could lead to important consequences for degravitation where the fundamental cosmological constant, here arising with tension of the 3-brane,  could be large but give rise only to a small backreaction on the geometry on timescales of the order of the graviton Compton wavelength.
As observational motivation, the model carries distinguishable signatures that could be observable at late times~\cite{niayeshghazal}.

In Cascading Gravity, our 3-brane is embedded in a succession of higher-dimensional branes, each
with their own intrinsic Einstein-Hilbert term. In the simplest realization, with 6D bulk space-time, the action is
\ba
\nonumber
S_{\rm 6D \; Cascading} &=& \frac{M_6^4}{2}\int_{\rm bulk} {\rm d}^6x \sqrt{-g_6} R_6 + \frac{M_5^3}{2}\int_{\rm 4-brane} {\rm d}^5x \sqrt{-g_5} R_5 \\
&+&   \int_{\rm 3-brane} {\rm d}^4x \sqrt{-g_4}\left(\frac{M_4^2}{2}R_4 + {\cal L}^{(4)}_{\rm matter}\right)\,.
\label{6Dcascade}
\ea
There are two characteristic crossover scales, $m_5 = M_5^3/M_4^2$ and $m_6 = M_6^4/M_5^3$. Assuming $m_6\ll m_5$, the gravitational potential
on our brane cascades from a $1/r$ (4D gravity) regime at short distances, to a $1/r^2$ (5D gravity) regime at intermediate distances, and
finally to a $1/r^3$ (6D gravity) regime at large distances. Remarkably, the codimension-1 kinetic term makes the 4D propagator
finite~\cite{cascade1,cascade2}, thereby regulating the logarithmic divergence characteristic of pure codimension-2 branes~\cite{cod2eft,massimo}.
Some of the cosmological implications of this model have been explored in~\cite{niayeshghazal,markwyman,nishant}.

The ghost issue is trickier. Perturbing around the flat space solution with empty branes, one finds that
a ghost scalar mode propagates. This is seen most directly from the tensor structure of the one-graviton exchange amplitude between conserved sources
in the UV limit:
\be
\mathcal A \; \sim \;
{\cal T}_4^{\mu\nu}\cdot  \frac{1}{k^2-i\varepsilon}\cdot \left({\cal T}'_{4 \mu\nu}-\frac
13 \eta_{\mu\nu}{\cal T}'_4\right)
- \frac{1}{6}\;{\cal T}_4\cdot \frac{1}{k^2-i\varepsilon}\cdot {\cal T}_4'\,.
\label{ampghost}
\ee
We have conveniently separated terms as the sum of a massive spin-2 contribution, with the well-known $1/3$ coefficient, plus a contribution from a conformally-coupled
scalar $\pi$. The problem is with this scalar mode. The last term in~(\ref{ampghost}) is negative, indicating
that $\pi$ is a ghost. This UV behavior is identical to other higher-dimensional scenarios~\cite{gigashif,sergei}.

However, it was immediately noticed~\cite{cascade1} that the ghost is removed if the codimension-2 is endowed with sufficiently large tension $\lambda$. (In analogy with a cosmic string in
ordinary 4D gravity, adding tension to a codimension-2 defect leaves the induced geometry flat but creates a deficit angle in the extra dimensions.) Instead of~(\ref{ampghost}) the
exchange amplitude now depends on the tension $\lambda$:
\be
\mathcal A \; \sim \;
{\cal T}_4^{\mu\nu}\cdot  \frac{1}{k^2-i\varepsilon}\cdot \left({\cal T}'_{4 \mu\nu}-\frac
13 \eta_{\mu\nu}{\cal T}'_4\right)
 +\frac{1}{6\left(\frac{3\lambda}{2m_6^2M_4^2}-1\right)}\;{\cal T}_4\cdot \frac{1}{k^2-i\varepsilon}\cdot {\cal T}_4'\,.
\label{ampnoghost}
\ee
The $\pi$ contribution is healthy, and the ghost is absent, provided that
\be
\lambda \geq  \frac{2}{3}m_6^2M_4^2\,.
\label{lambound}
\ee
Note the essential role of the codimension-1 brane. In the limit $M_5\rightarrow 0$ where its induced gravity term disappears, the lower
bound in~(\ref{lambound}) diverges. In particular, adding tension cannot cure the ghost in the pure codimension-2 DGP case~\cite{gigashif,sergei}.

A limitation of the calculation presented in~\cite{cascade1} is that~(\ref{ampnoghost}) and~(\ref{lambound}) were obtained through a decoupling limit of~(\ref{6Dcascade}),
namely $M_5,M_6\rightarrow \infty$ keeping $\Lambda_6 = (m_6^4M_5^3)^{1/7}$ fixed.
(The derivation of these results using the decoupling limit is reviewed in Appendix A.)
Although the decoupling limit in DGP has been shown to faithfully reproduce much of the phenomenology of the full-fledged
higher-dimensional theory~\cite{lpr,nathan}, a complete 6D calculation is clearly needed to establish unequivocally the consistency of the cascading gravity framework.
In particular, the analysis of~\cite{cascade1} could not demonstrate the absence of ghost modes that disappear in the decoupling limit.

In this paper, we study the issue of stability rigorously by perturbing the full 6D action~(\ref{6Dcascade}) around a background including tension
on the 3-brane. The background geometry is flat everywhere, but the extra dimensions show a deficit angle due to the codimension-2 source, with
the 3-brane located at the conical singularity.

We are immediately faced with a puzzle: brane tension, by virtue of being a source for gravity, cannot by itself modify the graviton kinetic term. Indeed, the
3-brane part of the quadratic lagrangian density takes the initial form
\be
S_{\rm 3-brane} = \int  \d ^4x \(-\frac {3\mq} 4\, \pi \Box_4
\pi- 2\pi^2 \lambda\)\,,
\label{L3ini}
\ee
where $\pi$ is the scalar perturbation of the 4D metric.
Thus $\pi$ acquires a localized mass term from $\lambda$, but as expected its kinetic term is unscathed and remains negative.
Incidentally, even the mass term is puzzling --- $\pi$ is a Goldstone mode and should enjoy a shift symmetry. How can~(\ref{L3ini})
be consistent with the conclusions of the decoupling analysis?

The resolution lies in the other scalar modes: integrating out the non-dynamical degrees of freedom in the bulk
results in contributions localized on the 3-brane, which generate $\lambda$-dependent additions to the $\pi$ kinetic term.
The final 3-brane lagrangian for $\pi$ is
\be
S_{\rm 3-brane} = \int  \d ^4x\, \frac{3M_4^2}{4}\left(\frac{3\lambda}{2m_6^2M_4^2}-1\right)\pi\Box_4\pi\,.
\ee
Provided that $\lambda \ge  2m_6^2M_4^2/3$, the kinetic term for $\pi$ is positive, in precise agreement with the decoupling limit.
Meanwhile, the puzzling mass term gets canceled by a bulk contribution. Our analysis
shows unambiguously that the cascading framework is ghost-free, at least perturbatively,
provided that the codimension-2 brane is endowed with sufficiently large tension. Whether the theory is
stable non-linearly remains of course an open issue that deserves further study.

The full 6D calculation we perform here is necessarily delicate because of the fact that the background geometry contains a conical singularity at the location of the 3-brane on which we intend to localize matter. This is the usual conical singularity associated with the tension of a codimension-2 defect.
The orbifold symmetry imposed across the codimension-1 brane creates a $\mathds{Z}_2$ reflection axis in the background conical solution which is further felt at the level of perturbations. Although there are many different ways of representing the background geometry, none is particularly straightforward for performing the perturbative analysis. To get a sense of the difficulty, consider the usual way of writing a conical defect,
\be
{\rm d}s^2={\rm d}r^2 + r^2(1-\delta)^2 {\rm d}\theta^2+\eta_{\mu\nu} {\rm d}x^{\mu}{\rm d}x^{\nu}\,.
\label{1stcoord}
\ee
The angular variable $\theta$ runs from 0 to $2\pi$, and $\delta$ is the deficit angle proportional to the tension.
The 3-brane is understood to be located at $r=0$. The problem with this coordinate system is that the orbifold brane is located on two disjoint surfaces: $\theta=0$ and $\theta=\pi$. Even if the geometry is smoothed out at $r=0$, it is clear that this metric will provide a poor description of physics on the codimension-1 brane near the codimension-2 brane.

Alternatively, the geometry can be put in the form
\be
{\rm d}s^2=e^{\phi(y,z)}\left({\rm d}y^2+{\rm d}z^2\right)+\eta_{\mu\nu} {\rm d}x^{\mu}{\rm d}x^{\nu}\,, \quad\,\,\, \text{with  } \,\, \phi(y,z)=-\delta \ln (y^2+z^2)\,.
\label{2ndcoord}
\ee
One advantage of this coordinate system is that the codimension-1 brane is now at $z=0$. The drawback is clear, however:
the conformal factor $\phi$ is singular at the location of the codimension-2 brane. One can deal with this by regulating the conformal factor, {\it e.g.}, $\phi \rightarrow -\delta \ln (y^2+z^2+\varepsilon^2)$, but this need to regulate the background geometry comes at the price of introducing extra terms in the perturbation equations which may or may not be negligible
in the limit $\varepsilon \rightarrow 0$. Although the final equations shall be independent of $\varepsilon$, it is necessary to introduce the artifice of regularization to obtain them.

Instead we will use a different form of the background metric, one that we have found is best suited for the analysis of perturbations:
\be
\d s^2=\left(1+\beta^2 \left(1-\epsilon(y)^2\right)\right)\d z^2+\left(\d y+\beta \epsilon(y) \d z\right)^2 +\eta_{\mu\nu} {\rm d}x^{\mu}{\rm d}x^{\nu}\,.
\label{ourcoord}
\ee
In this coordinate system, the codimension-1 brane is still at $z=0$, and we work in the half-picture $z> 0$. The above metric is understood to be orbifolded at $z=0$. The parameter
$\beta$ is related to the 3-brane tension. Meanwhile, $\epsilon(y)$ is a regulating function with the following properties: $\epsilon(\infty)=1$, $\epsilon(-y)=-\epsilon(y)$
and $\epsilon'(y)=2\delta_{\epsilon}(y)$, where $\delta_{\epsilon}(y)$ is an explicit regularization of the Dirac delta function. The upshot of this coordinate system is that
the induced metric on the codimension-1 brane is simply Minkowski space-time, and similarly for the induced geometry on the codimension-2 brane at $z=y=0$.
As we will see below, this greatly simplifies the analysis of perturbations.

As always the presence of ghosts in a theory can be determined by checking the sign of the kinetic term of all propagating degrees of freedom in the action. In the case of gravitational theories, however, the inevitable presence of gauge modes and constraints complicates matters. Only after the gauge has been completely fixed and the constraints fully solved will the action reduce to the true physically propagating degrees of freedom whose kinetic terms can be inspected.
As is well known, before the constraints have been solved the action may well contain wrong-sign kinetic terms, {\it e.g.} the famous conformal factor problem of Euclidean quantum gravity.

One way to bypass this problem is to compute the coupling to a conserved source. Since gauge invariance enforces conservation of stress energy, any gauge degrees of freedom will not couple to a conserved stress-energy source, and, in particular, will not contribute to the single-graviton exchange amplitude. In the following we shall do precisely this to verify the absence of ghost degrees of freedom.

The paper is organized as follows. In Sec.~\ref{background}, we describe the background geometry~(\ref{ourcoord}), starting from the general ADM form of
the 6D metric. In Sec.~\ref{pertn1}, we study metric perturbations, focusing first on 4D vector and tensor modes. We then turn to 4D scalar perturbations in
Sec.~\ref{pertn2}, and derive the lower bound on the tension. In Sec.~\ref{externalcoupling}, we discuss couplings to external sources.
We conclude with a summary of the results and discuss future avenues in Sec.~\ref{conclu}. The Appendices include a review of the decoupling
calculation (Sec.~\ref{decoup}), an argument for why certain naively ghostly terms do not contribute to the imaginary part of the exchange amplitude (Sec.~\ref{sigma1}), and an argument in (Sec.~\ref{sigma2}) for why one of the scalar modes,
which at first sight looks problematic and ghostly, is in fact pure gauge and does not contribute to the exchange amplitude. Our notation throughout is as follows: $A,B,\ldots$ denote 6D space-time indices, $a,b\ldots$ are 5D indices, and $\mu,\nu,\ldots$ are 4D indices. The coordinates along the worldvolume of the 3-brane are denoted
 by $x^\mu$, whereas the extra dimensions have coordinates $y$ and $z$. The 4-brane is located at $z=0$, and the 3-brane at $y=z=0$.

\section{Cascading Setup}
\label{background}

Our approach is to start with the full 6D action for the cascading setup written in the radial ADM formalism, and then perturb this action to quadratic order.
We work in the ``half (or reduced)-picture", where we restrict ourselves to the $z>0$ side of the $\mathds{Z}_2$-symmetric 4-brane.
Crucially this implies that all quantities evaluated at $z=0$ are understood in the following sense: $\phi(z=0)=\lim_{z \rightarrow 0^+} \phi(z)$.

In the ADM form, with $z$ playing the role of a ``time" variable, the 6D metric in the region $z>0$ is
\be
\label{6Dbackground}
\d s^2= \gamma_{AB}\d x^A \d x^B=N^2 \d z^2+g_{ab}(\d x^a+ N^a \d z)(\d x^b+N^b \d z)\,.
\ee
In what follows, covariant objects are constructed using the 5D metric $g_{ab}$ induced on constant-$z$ surfaces.
The 6D part of the action, including the appropriate Gibbons-Hawking boundary term, is then
\ba
S_6=\frac{M_6^4}{2}\int^{+}\hspace{-5pt}\d z \int \d^5x \, \mathcal{L}_6\,,
\ea
with $\int^+ \d z=\int_0^{+\infty} \d z$, and where
\ba
\mathcal{L}_6\equiv \sqrt{-g_6}\, \( R^{(5)}+ {K^a_{\;a}}^2-K^a_{\;b} K^b_{\;a}\)\,.
\ea
The full action is given by
\ba
\nonumber
S &=& \int^{+} \d^6 x \left( \frac{M_6^4}{2} \mathcal{L}_6 +  {\mathcal L}_{\rm matter}\right)+\frac{M_5^3}{2} \int_{z=0} \d^5 x \sqrt{-g_5} R_5 \\
& & + \int_{z=y=0} \d^4 x \sqrt{-g_4}\(\frac{M_4^2}{2} R_4-\lambda \),
\ea
where $\mathcal{L}_{\rm matter}$ is the Lagrangian for the external sources which may be distributed in 6D, or localized on either brane. In what follows we will compute the coupling for an arbitrary conserved 6D, 5D or 4D source.

\subsection{Background Solution}
{\label{backgroundsolution}}

There are several ways to write the background solution. However, as discussed earlier,
great care must be taken because of the singular nature of the geometry. Consider the following background metric
\be
\d s^2=\bar \gamma_{AB}\d x^A \d x^B=(1+\beta^2 (1-\epsilon(y)^2))\d z^2+(\d y+\beta \epsilon(y) \d z)^2 + \eta_{\mu\nu} {\rm d}x^{\mu}{\rm d}x^{\nu}\,.
\label{ourcoord2}
\ee
By comparing with~(\ref{6Dbackground}), we can read off the background expressions for the ADM variables:
\ba
\nonumber
\bN^{2}&=&1+\beta^2(1-\epsilon(y)^2)\;; \\
\nonumber
\bN^{y}&=& \beta \epsilon(y)\;; \qquad \bN^{\;\mu}= 0\;; \\
\overline{g}_{ab}&=&\eta_{ab}\,.
\ea
The regulating function $\epsilon(y)$ is such that $\epsilon(\infty)=1$, $\epsilon(-y)=-\epsilon(y)$ and $\epsilon'(y)=2\delta_{\epsilon}(y)$, where $\delta_{\epsilon}(y)$ is an explicit regularization of the delta function. The induced metrics on the codimension-1 ($z=0$) and codimension-2 ($z=y=0$) branes are both flat Minkowski space.
The 6D Riemann tensor vanishes on the background for $z>0$, as it should. For instance, since $\bN^{2}+\bN_{y}^{2}=1+\beta^2$, the Ricci scalar
is given by
\ba
\bar{R}_6=-\frac{1}{\bN} \py\(\frac{ \py \left(\bN^{2}+\bN_{y}^{2}\right)}{\bN}\)=0\,.
\ea
Because this metric is orbifolded at $z=0$, it follows from the Isra\"el junction conditions that~(\ref{ourcoord2}) correctly describes the metric of a codimension-2 object with tension localized on the orbifold plane. Explicitly, the Isra\"el junction conditions are
\be
M_6^4 (K \delta^a_{\;b} - K^a_{\;b})\Big|_{z=0}=T^a_{\;b}-M_5^3 G^a_{\;b}\,.
\label{Israel1}
\ee
In the background, we have $T^a_{\;b}=-\lambda\delta(y)$ and $G^a_{\;b}=0$. Meanwhile,
in the gauge \eqref{6Dbackground}, the extrinsic curvature on the surface $z=0$ is
\ba
K_{ab}&=&\frac{1}{2N} \(\partial_z g_{ab}-\nabla_{a} N_b-\nabla_b N_a \) \nn \\
&=& \frac{1}{2N} \( \partial_z g_{a b}-g_{a c} \partial_{b}N^{c}-g_{b c} \partial_{a}N^{c}-N^{c}\partial_{c} g_{a b}\)\nn\\
&=&-\frac{\py \bN_{y}}{\bN} \delta^y_{\; a} \delta^y_{\; b}\,.
\ea
At the background level, the Isra\"el matching conditions~(\ref{Israel1}) therefore read
\ba
\label{Israel}
K_{yy}=-\frac{\py \bN_{y}}{\bN} =-\frac{\beta \epsilon'(y)}{\sqrt{1+\beta^2(1-\epsilon^2(y))}}=-\frac{\lambda}{\ms}\delta(y)\,.
\ea
Since $\epsilon'(y) = 2\delta_\epsilon(y)\rightarrow \delta(y)$, the extrinsic curvature does indeed encode the codimension-2 source.
The relation between the brane tension $\lambda$ and the required value for $\beta$ requires some care, but can be determined
unambiguously by integrating the extrinsic curvature,
\ba
\int _{-\infty}^{+\infty}\hspace{-6pt}\d y  \frac{\py \Ny}{\bN} =\int_{-\infty}^{+\infty}\hspace{-6pt}\d y  \frac{\beta \epsilon'(y)}{\sqrt{1+\beta^2(1-\epsilon^2(y))}}
=\int_{-1}^{1}\frac{\beta \, \d \epsilon(y)}{\sqrt{1+\beta^2(1-\epsilon^2(y))}}=2\arctan \beta \,.
\ea
Thus the relation is
\ba
\lambda = 2 \ms\arctan \beta\,.
\ea
In particular, we recover the standard result that this static solution can only hold a maximal positive codimension-2 tension: $\lambda< \pi \ms$. (This is one-half of the usual condition because the $\mathds{Z}_2$ orbifolding projects out half of the space.)

\section{Tensor and Vector Perturbations}
\label{pertn1}

To study perturbations, we perform a scalar-vector-tensor decomposition of the full 6D metric $\gamma_{AB}=\bar \gamma_{AB}+\delta \gamma_{AB}$
 with respect to the 4D Lorentz group. The tensor and vector modes are straightforward as none of them couple to the positions of the branes and are given by a simple generalization of the massless scalar field. In particular, the tensor and vector sectors are manifestly ghost free, regardless of the value of the tension on the codimension-2 brane.

\paragraph*{Tensors:}
The metric for the tensor perturbations takes the form
\ba
\nonumber
\delta \gamma_{AB}\d x^A \d x^B &=& h^{\rm T}_{\mu \nu}\d x^\mu\d x^\nu \nn \,,
\ea
where $\eta^{\mu\nu} h^{\rm T}_{\mu\nu}=0$ and $\partial^{\mu} h^{\rm T}_{\mu\nu}=0$.
The action for the tensor perturbations is given by
\ba
S&=&\frac{M_6^4}{8} \int^+ \d z \int \d^5 x \( -\bar N \partial_a h_{\rm T}^{\mu \nu} \partial^a h^{\rm T}_{\mu\nu}-\frac{1}{\bar N}\( \Ln h_{\rm T}^{\mu\nu} \)^2 \)\,, \\
&  +&\frac{M_5^3}{8} \int_{z=0} \d^5 x \( h_{\rm T}^{\mu \nu} \Box_5 h^{\rm T}_{\mu\nu}\)+\frac{M_5^3}{8} \int_{z=y=0} \d^4 x \( h_{\rm T}^{\mu \nu} \Box_4 h^{\rm T}_{\mu\nu} \) \, ,\nn
\ea
where $\Ln$ is proportional to the Lie derivative: $\Ln X=\partial_z X-\Ny \py X$. Notice that $\Ln$ and $\partial_y$ do not commute:
\ba
\py \Ln X=\Ln \py X-\py \Ny \, \py X\,.
\label{Lniden}
\ea
The tensor modes constitute 5 degrees of freedom, each of which satisfies the same equation which is manifestly ghost free since the action is only composed of a sum of positive kinetic terms. These five spin-2-degrees of freedom can be further decomposed into two helicity-2, two helicity-1 and one helicity-0 modes. The helicity-0 degree of freedom is expected to exhibit the Vainshtein mechanism as in usual massive gravity~\cite{vainshtein,nima,babichev}, DGP model~\cite{lpr,ddgv,gruzinov,lue,greg} and general theories of resonance graviton~\cite{degrav,gd}.

\paragraph*{Vectors:}
For the vector perturbations the metric takes the form
\ba
\delta \gamma_{AB}\d x^A \d x^B &=& 2 \( A_{\mu}+\bar N_y B_{\mu}\) \d x^{\mu}\d z+2 B_{\mu} \d x^{\mu}\d y +\( \partial_{\mu} C_{\nu}+ \partial_{\nu} C_{\mu}\)\d x^\mu\d x^\nu \,,
\ea
where the vectors $A_{\mu},B_{\mu}$ and $C_{\mu}$ are all transverse: $\partial_{\mu}A^{\mu}=\partial_{\mu}B^{\mu}=\partial_{\mu}C^{\mu}=0$. There is sufficient gauge freedom in the vector sector to set $B_{\mu}=0$. The action for the remaining perturbations is then given by
\ba
S&=&\frac{M_6^4}{8} \int^+ \d^6x \( -\frac{2}{\bar N}\( \partial_y {A}_{\mu}\)^2+\frac{2}{\bar N} \(A_{\mu}-\Ln C_{\mu}\) \Box_4 \(A^{\mu}-\Ln C^{\mu}\)+2\bar N \partial_y C^{\mu}\Box_4 \partial_y C_{\mu} \) \nn \\
&&+\frac{M_5^3}{8} \int_{z=0} \d^5 x \(2 \partial_y C^{\mu}\Box_4 \partial_y C_{\mu}\) +\int^+ \d^6x \, \bN \(A_\mu T^{\mu z}-C_\mu \partial_\nu T^{\mu\nu}\) \, ,
\ea
where $T^{AB}$ is an external 6D source, (which can include localized contributions on either of the branes).
Note that neither vector field has a 4D source and so we expect that the vectors will be appropriately smooth at $y=0$. With this in mind, we may neglect the mild $y$-dependence of $\bar N$, since its departure from $\bar N =1$ will be of zero measure in the integral. Then, on performing the following local field redefinition,
\be
\hat{C}_{\mu}=\partial_y C_{\mu}\,; \, \quad \hat{A}_{\mu}=A_{\mu}-\Ln C_{\mu}\,,
\ee
the kinetic terms are diagonal and manifestly positive definite and so is the coupling to the source
\ba
S_{\rm kin}&=&\frac{M_6^4}{8} \int^+ \d^6x \( \frac{2}{\bar N} \hat A_\mu \Box_4 \hat A^\mu+2\bar N \hat C^{\mu}\Box_4 \hat C_{\mu} \)
+\frac{M_5^3}{8} \int_{z=0} \d^5 x \(2 \hat C^{\mu}\Box_4 \hat C_{\mu}\)  \\
S_{\rm sources}&=&-\int^+ \d^6x\, \bN \hat C_\mu \(T^{\mu y}+\Ny T^{\mu z}\) \, ,
\ea
where we used conservation of energy along the transverse direction $\nabla_A T^{A\mu}=0$.

Notice that $\hat{A}_{\mu}$ lives entirely in 6D, whereas $\hat{C}_{\mu}$ has a kinetic term both in 6D and 5D. Crucially all the kinetic terms are positive, as required, confirming that there are no ghosts present in the vector sector.

\section{Scalar Perturbations}
\label{pertn2}

The scalar perturbations are tricky because the positions of the brane transform as 4D scalars. We choose to utilize the bulk gauge freedom to work in a gauge where the codimension-1 and -2 branes remain at fixed position, respectively $z=0$ and $z=y=0$.

At the perturbed level, the metric has seven 4D scalar modes:
\ba
\nonumber
\delta \gamma_{AB}\d x^A \d x^B &=& h_{zz}\d z^2+2 h_{zy}\d z \d y + h_{yy}\d
y^2+2 h_{\mu z}\d x^\mu\d z+ 2 h_{\mu y}\d x^\mu\d y \\
&+&\(\pi \eta\mn +\partial_\mu \partial_\nu \varpi\)\d x^\mu\d x^\nu \nn \,.
\ea
One can set one of these modes to zero by fixing the gauge while
keeping the brane positions unperturbed. We use this freedom to set
$\varpi=0$, such that $h_{\mu\nu} = \pi \eta_{\mu\nu}$. This is an analogue
of conformal Newtonian/longitudinal gauge used in cosmological perturbation theory.
We are therefore left with six scalar degrees of freedom, which will be
denoted by $\chi, \phi, V, \sigma, \tau$ and $\pi$:
\ba
\nonumber
h_{ab}\d x^a \d x^b & =& \pi \eta_{ab}\d x^a \d x^b+V \d y^2+2
\partial_\mu \tau\  \d x^\mu \d y\\
\nonumber
h_{zz}&=& 2\bN^{\;2} \phi+ \bN^{a} \bN^{b} h_{ab}+2\bN^{a} A^b
\eta_{ab}\\
h_{az}&=&A^b\eta_{ab}+\bN^{b} h_{ab}\,,
\ea
where $A_a\d x^a = \partial_a \sigma \d x^a+\chi \d y$.
Note that $A^a$ is the perturbation in the shift vector, $N^a = \bN^a + A^a$, while
$\phi$ is the perturbation is the lapse function, $N = \bN(1+\phi)$.

To quadratic order in perturbations, the full action is given by
\ba
\label{Linitial}
\nonumber
S_{\rm quad} &=&  \frac{\ms}{2} \int^+ \d^6x  \mathcal{L}_6
- \frac {3 \mf} 4 \int_{z=0} \d^5x \, \pi \(\Box_4 V+2 \Box_5 \pi-2 \Box_4\py
\tau\) \\
&-&\int_{z=y=0} \d^4x \(\frac {3\mq} 4\, \pi \Box_4
\pi+ \lambda \pi^2 \)\,.
\ea
The kinetic term for $\pi$ on the codimension-2 brane has manifestly the wrong sign. However, as we will see,
the presence of brane tension $\lambda$ can change the sign of this kinetic term, after integrating
out the non-dynamical degrees of freedom. Meanwhile, the mass term $\lambda\pi^2$ localized on the codimension-2 brane,
which seemingly breaks the shift symmetry, will cancel out after integrating by parts a bulk contribution.

To see how this works in detail, let us expand the 6D action to quadratic order, still focusing on the region $z>0$.
It is convenient to perform the following field redefinitions
\ba
\tilde\chi&=&\chi-(\Ln +\py \Ny)\tau-2\frac{\py \bN}{\bN}\ts + \frac{4}{\Box_4} \Ln \py \pi\\
\Box_4 \tilde \sigma &=& \Box_4 \sigma-2 \Ln\pi=-\bN \delta K^\mu_{\; \mu}\\
\Box_4 \tilde \tau &=& \Box_4 \tau-2 \py \pi\,.
\ea
In terms of these new field variables, the result for ${\cal L}_6$ at quadratic order is then
\ba
\label{6Dcurvature}
\mathcal{L}_6&=&\frac {1}{2\bN} \tilde\chi\Box_4\tilde\chi
-\bN \( V+4\pi -2  \py \tilde \tau+2 \frac{\py \Ny}{\bN^2}\tilde \sigma\)\Box_4 \phi \nn \\
&&-\frac {1}{\bN} \py \Ny \Ln (\tilde \tau \Box_4 \tilde \tau)
+3 \bN \pi (\py^2-\Box_4) \pi-\frac{3}{\overline N} (\Ln \pi)^2
-2\frac{\py \Ny}{\bN}\partial_z \pi^2\nn\\
&&
-\frac{1}{\bN}\Box_4 \ts (\Ln - \py \Ny) (V+4\pi)-\frac{3}{\bN} \py \Ny \pi \Box_4 \ts
+\frac{2}{\bN} \Box_4 \ts \py \Ln \tilde \tau
+3 \bN \pi \Box_4 \py \tilde \tau\nn\\
&&
-\frac 3 2 \bN \pi \Box_4 V
+2 \frac{\py \Ny}{\bN}\frac{\py \bN}{\bN} \ts \Box_4 \tilde \tau-\frac{\py^2 \bN}{\bN}\ts \Box_4 \ts
+\py \bN \tilde \tau \Box_4 (V+4\pi) \,.
\ea
The issue of the $\pi$ mass term in~(\ref{Linitial}) can be immediately resolved. The last term in
the second line of~\eqref{6Dcurvature}, when integrated by parts, generates a mass term for $\pi$
on the 3-brane that precisely cancels that in~\eqref{Linitial}:
\ba
-2\ms\int^+\hspace{-3pt}\d z \int \d y\frac{\py \Ny}{\bN}\partial_z \pi^2&=&2\ms\int^+ \hspace{-3pt} \d z \int \d y \(\partial_z \frac{\py \Ny}{\bN}\) \pi^2
-2\ms\int \d y \Big[\frac{\py \Ny}{\bN} \pi^2  \Big]_{z=0}^{\infty}\nn\\
&=&2\ms\int_{z=0} \d y \frac{\py \Ny}{\bN} \pi^2=\lambda \pi^2\,.
\ea
This cancelation makes manifest the shift symmetry of $\pi$. We now proceed to decipher the $\pi$ kinetic term,
which requires a few field redefinitions and integrating out non-dynamical fields.

\subsection{Lagrange multipliers $\phi$ and $V$}

As it stands,~(\ref{6Dcurvature}) is linear in $\phi$ and $V$.
Varying the action with respect to either of these therefore yields a constraint
on the other degrees of freedom. We choose to vary with respect to $\phi$
and interpret the result as a constraint for $V$. Upon substituting $V$
by its constrained value back into the action, we will see that the action picks up a quadratic term in $\tilde \sigma$.

The expression for $V$ that follows from the $\phi$ variation is
\ba
\label{EqforV}
V=-4 \pi+ 2 \py \tilde \tau-\frac{2\py \Ny}{\bN^2} \ts\,.
\ea
In the presence of an external source, there is an additional $T_6^{zz}$ that enters in this equation;
we shall take this into account in Sec.~\ref{externalcoupling} when we compute the relevant couplings  --- see Eq.~(\ref{newV}).
Substituting this expression for $V$ in the 6D action, we can simplify the last two lines in~\eqref{6Dcurvature} as follows
\ba
\nonumber
A&\equiv & -\frac{1}{\bN}\Box_4 \ts (\Ln - \py \Ny) (V+4\pi)-\frac{3}{\bN} \py \Ny \pi \Box_4 \ts
+\frac{2}{\bN} \Box_4 \ts \py \Ln \tilde \tau
+3 \bN \pi \Box_4 \py \tilde \tau\nn\\
&&
-\frac 3 2 \bN \pi \Box_4 V
+2 \frac{\py \Ny}{\bN}\frac{\py \bN}{\bN} \ts \Box_4 \tilde \tau-\frac{\py^2 \bN}{\bN}\ts \Box_4 \ts
+\py \bN \tilde \tau \Box_4 (V+4\pi) \nn\\
&=& -\frac 1 \bN\Box_4 \ts (\Ln - \py \Ny) \(2 \py \tilde \tau-\frac{2\py \Ny}{\bN^2} \ts\)-\frac 3 \bN\py \Ny \pi \Box_4 \ts
+\frac 2 \bN \Box_4 \ts \py \Ln \tilde \tau \nn
 \\
&&+3 \bN \pi \Box_4 \py \tilde \tau- \frac 3 2 \bN \pi \Box_4 \(-4 \pi+ 2 \py \tilde \tau-\frac{2\py \Ny}{\bN^2} \ts\)
+2\frac{\py \Ny}{\bN}\frac{\py \bN}{\bN}\ts \Box_4 \tilde \tau\nn\\
&&-\frac{\py^2 \bN}{\bN^2}\ts \Box_4 \ts
+\py \bN \tilde \tau \Box_4 \(2 \py \tilde \tau-2 \frac{\py \Ny}{\bN^2}\ts\) \nn \\
&=& 6\bN \pi \Box_4 \pi-\py^2 \bN \tilde \tau\Box_4 \tilde \tau+\frac 2 \bN \Box_4 \ts  (\Ln-\py \Ny) \(\frac{\py \Ny}{\bN^2} \ts\)-\frac{\py^2\bN}{\bN}\ts \Box_4 \ts\,,
\ea
where in the last step we have used $(\Ln-\py \Ny)\py \tilde \tau=\py(\Ln \tilde \tau)$ from~(\ref{Lniden}).
We can further simplify this expression by integrating by parts:
\ba
\int^+ \hspace{-5pt}\d^6x A &=&\int^+ \hspace{-5pt}\d^6x\(6\bN \pi \Box_4 \pi-\py^2 \bN \tilde \tau\Box_4 \tilde \tau+ \frac 2 \bN \Box_4 \ts  (\Ln-\py \Ny) \(\frac{\py \Ny}{\bN^2} \ts\)
-\frac{\py^2\bN}{\bN}\ts \Box_4 \ts\)\nn\\
&=&\int^+ \hspace{-5pt}\d^6x \(6\bN \pi \Box_4 \pi-\py^2 \bN \tilde \tau\Box_4 \tilde \tau - \frac {\ts \Box_4 \ts}{2\bN^2}\, \py\left[\frac{\py (\bN^2+\Ny^2)}{\bN}\right]\) - \int_{z=0} \hspace{-5pt}\d^5x \frac{\py \Ny}{ \bN^3}\ts \Box_4 \ts\nn \\
&=&\int^+ \hspace{-5pt}\d^6x \(6\bN \pi \Box_4 \pi-\py^2 \bN \tilde \tau\Box_4 \tilde \tau\)- \int_{z=0} \d^5x \frac{\py \Ny}{ \bN^3}\ts \Box_4 \ts \,,
\label{Asimp}
\ea
where the last follows because $\bN^2+\Ny^2=1+\beta^2 = {\rm constant}$.
Substituting~(\ref{Asimp}) back into~(\ref{6Dcurvature}), we note that the quadratic terms in $\tilde \tau$ simplify through integration by parts:
\ba
-\int^+\hspace{-5pt}\d z \, \(\py^2 \bN+\frac{\py \Ny}{\bN}\Ln  \)(\tilde \tau \Box_4 \tilde \tau)
&=&-\frac 12 \int^+\hspace{-5pt}\d z \, \py \(\frac{\py(\bN^2+\Ny^2)}{\bN}\)\tilde \tau \Box_4 \tilde \tau
-\left[\frac{\py \Ny}{\bN}\tilde \tau \Box_4 \tilde \tau\right]_0^{\infty} \nn\\
&=&\int_{z=0} \d^5x \frac{\py \Ny}{\bN} \tilde \tau \Box_4 \tilde \tau \,.
\ea
Combining all of these results, the 6D part of the quadratic action reduces to
\ba
\label{6Daction2}
\int^+ \d^6 x \mathcal{L}_6&=&\int^+ \d z \d^5 x \(\frac {1}{2\bN}\tilde\chi\Box_4\tilde\chi
+3 \bN \pi \Box_5 \pi-\frac 3\bN (\Ln \pi)^2\) \nn \\
&+&\int_{z=0} \d^5x \frac{\py \Ny}{\bN}\(\tilde \tau \Box_4 \tilde \tau-\frac{1}{\bN^2}\ts \Box_4 \ts\)\,.
\ea

Similarly, we must also substitute~(\ref{EqforV}) for $V$ into the 5D part of~(\ref{Linitial}):
\ba
\label{action5}
\mathcal{L}_5= \left.\frac 12 \sqrt{-g_5}R_5\right\vert_{\rm quadratic}=\frac 32 \pi \Box_5 \pi +\frac 32 \frac{\py \Ny}{\bN^2} \pi \Box_4 \ts \,.
\ea
As we will see, the above kinetic mixing term between $\pi$ and $\ts$ is what provides an additional 4D kinetic term for $\pi$.

Putting everything together, and using the background Isra\"el Matching condition \eqref{Israel}, $\ms \py \Ny/\bN = \lambda\delta(y)$,
the complete action for the scalar modes is given by
\ba
S_{\rm quad} &=&\frac{\ms}{2} \int^+ \d^6 x  \left[ \frac {1}{2\bN}\tilde\chi\Box_4\tilde\chi
+3\(\bN \pi \Box_5 \pi-\frac 1\bN (\Ln \pi)^2\) \right] \nn \\
&+&\mf \int_{z=0} \d^5x\, \frac 32 \pi \Box_5 \pi\nn \\
&+& \mq \int_{z=y=0} \d^4 x \left[-\frac{3}{4}\pi \Box_4 \pi +
\frac{\lambda}{2\mq}
\(\tilde \tau \Box_4 \tilde \tau-\frac{1}{\bN^2}\ts \Box_4 \ts+\frac{3}{m_6 \bN}\pi \Box_4 \ts\)\right]\,.
\ea

\subsection{$\ts$ degree of freedom}

The last step consists of diagonalizing the kinetic matrix for $\pi$ and $\ts$. This is achieved by the following
change of variable
\ba
\label{sigmabar}
\hat \sigma=\tilde \sigma - \frac{2 \bN}{4 m_6}\,  \pi\,,
\ea
in terms of which the action becomes
\ba
\label{action3}
S_{\rm quad} &=&\frac{\ms}{2} \int^+ \d^6 x  \left[ \frac {1}{2\bN}\tilde\chi\Box_4\tilde\chi
+3\(\bN \pi \Box_5 \pi-\frac 1\bN (\Ln \pi)^2\) \right] \nn \\
&+&\mf \int_{z=0} \d^5x\, \frac 32 \pi \Box_5 \pi\nn \\
&+&\mq \int_{z=y=0}  \d^4 x \left[ \frac 3 4 \(\frac{3\lambda}{2m_6^2\mq}-1\) \pi\Box_4 \pi +
\frac{\lambda}{2\mq}
\(\tilde \tau \Box_4 \tilde \tau-\frac{1}{\bN^2}\hat\sigma \Box_4 \hat\sigma\) \right] \,.
\ea
This is our main result. Through a series of field redefinitions, and after integrating out auxiliary fields, the resulting
kinetic term for $\pi$ on the codimension-2 brane acquires a $\lambda$-dependent contribution:
\ba
\mathcal L_2^{{\rm kin}}=\frac{3\mq}{4} \(\frac {3\lambda}{2m_6^2\mq}-1\)\pi \Box_4
\pi\,,
\ea
Thus the $\pi$ mode is healthy as long as the tension is larger than the lower bound
\ba
\lambda_{\rm min}=\frac{2}{3}\mq m_6^2\,.
\ea

\section{Coupling to an external source}
\label{externalcoupling}

It is immediately apparent from~(\ref{action3}) that $\pi$, $\tilde\chi$ and $\tilde \tau$ are manifestly unitary. However, $\hat \sigma$ appears as a ghost and would seem to present a serious concern. (This issue never arose in~\cite{cascade1} since this mode disappears in the decoupling limit.) As already discussed, though, it is common in the analysis of gravitational systems to find apparently ghostly degrees of freedom that nevertheless decouple when a physical, gauge-invariant calculation is performed.
In this case we will see that neither $\hat \sigma$ nor $\tilde \tau$ couple to conserved matter, and consequently these modes are pure gauge, at least to this order in perturbation theory. To see this, let us introduce the matter content in the action
\ba
\int^+ \d z \int \d ^5x \, \mathcal{L}_{\rm matter}=\int^+ \d z\int \d ^5 x \, \frac{\bar N}{2} h_{AB} T^{AB}_6\,,
\ea
where we consider a 6D stress-energy tensor in all generality. In particular the total stress-energy $T^{AB}_6$ can include terms localized on each of the branes:
\ba
T^{AB}_6=\mathcal{T}^{AB}_6+\bar N^{-1}\delta_+(z)\, \mathcal{T}^{ab}_5 \delta_a^A\delta_b^B+
\bar N^{-1}\delta(y)\delta_+(z)\, \mathcal{T}^{\mu\nu}_4 \delta_\mu^A\delta_\nu^B\,,
\ea
where $\delta_+(z)$ is the delta function normalized on the half-line: $\int_0^\infty \d z \,\delta_+(z) =1$. However, it is not necessary for us to distinguish between these different contributions since, in any case, any source must be covariantly conserved in a 6-dimensional sense. In Appendix B we discuss a subtle issue having to do with conservation of 5-dimensional external sources of the above form that appears to arise because of the singular nature of the background geometry.

In the presence of an external source, the constraint equation obtained by varying with respect to $\phi$ gets modified from~\eqref{EqforV} to
\ba
\label{newV}
V=-4 \pi+ 2 \py \tilde \tau-\frac{2\py \Ny}{\bN^2}\ts+\frac{\bar N^2}{\Box_4+i \varepsilon}T^{zz}_6 \,.
\ea
On substituting this back into the action, everything that is linear in $\phi$ drops out, including part of the matter action.
The remaining matter contribution which sources~(\ref{action3}) is
\ba
 \mathcal{L}'_{\rm matter}&=&-\bN \hat \sigma \Big[\partial_A T^{Az}_6+\frac{\py \Ny}{\bN^2}(T^{yy}_6-\Ny T^{yz}_6)\Big] \nn \\
&&-\bN \tilde \tau\Big[\partial_A T^{Ay}_6+\Ny \partial_A T^{Az}_6- \frac{\Ny \py \Ny}{\bN^2} (T^{yy}_6+\Ny T^{yz}_6)\Big]
+\bN \tilde \chi \Big[T^{yz}_6+\Ny T^{zz}_6\Big]\nn\\
&&-\frac{3\bN^2}{2m_6} \pi \Big[\partial_A T^{Az}_6+\frac{\py \Ny}{\bN^2}(T^{yy}_6-\Ny T^{yz}_6)\Big]
+\bN \pi \(2T^{\;\mu}_{6\; \mu}-\frac{3}{2} \bar \gamma_{AB}T^{AB}_6\)\nn\\
&&+2\bN \frac{\pi}{\Box_4}\Big[\partial_\mu \partial_z T^{\mu z}_6+\partial_\mu \partial_y T^{\mu y}_6-\frac{\Ny \py \Ny}{\bN^2}\partial_\mu T^{\mu y}_6\Big] \nn\\
&&+\frac{\bar N^3}{M_6^4} T^{zz}_6 \frac{1}{\Box_4+i \varepsilon} \( T^{yy}_6+2\bar N_y T^{yz}_6+\bar N_y^2 T^{zz}_6\) \nn \\
&=&-\bN \hat \sigma \nabla^{(6)}_A T^{Az}_6-\bN \tilde \tau \Big[\nabla^{(6)}_A T^{Ay}_6+\Ny \nabla^{(6)}_A T^{Az}_6\Big]
+\bN \tilde \chi  \Big[T^{yz}_6+\Ny T^{zz}_6\Big] \nn\\
&&-\frac{3\bN^2}{2m_6}\pi \nabla^{(6)}_A T^{Az}_6
+2\bN\frac{\pi}{\Box_4} \partial_\mu \nabla^{(6)}_A T^{A\mu}_6
+\bN\pi\Big[2T^{\;\mu}_{6\; \mu}-\frac{3}{2} \bar \gamma_{AB}T^{AB}_6-2\frac{\partial_\mu\partial_\nu}{\Box_4}T^{\mu\nu}_6\Big]   \nn \\
&&+\frac{\bar N^3}{M_6^4} T^{zz}_6 \frac{1}{\Box_4+i \varepsilon} \( T^{yy}_6+2\bar N_y T^{yz}_6+\bar N_y^2 T^{zz}_6\)\,,
\ea
where $T^{\;\mu}_{6\; \mu}=\eta_{\mu\nu} T^{\mu\nu}_6$, and $\nabla^{(6)}_A$ is the 6D covariant derivative.
In performing several integration by parts we have made the assumption that $\left.T^{zA}_6\right\vert_{z=0}=0$, which ensures that there is no external force on the orbifold plane. This condition is only necessary in the `half-picture' we are using here. In the doubled picture these terms are canceled by identical contributions of opposite sign on the other side of the orbifold plane.

We see that both modes $\hat \sigma$ and $\tilde \tau$ couple only to the transverse part of the stress-energy tensor. For conserved matter, $\nabla^{(6)}_A T^{AB}_6=0$, both $\hat \sigma$ and $\tilde \tau$ decouple, and the matter sector only sources $\tilde \chi$ and $\pi$
\ba
\mathcal{L}'_{\rm matter}&=&\bar N \tilde\chi  \Big[T^{yz}_6+\Ny T^{zz}_6\Big]
+\bar N \pi\Big[2T^{\;\mu}_{6\; \mu}-\frac{3}{2} \bar \gamma_{AB}T^{AB}_6-2\frac{\partial_\mu\partial_\nu}{\Box_4}T^{\mu\nu}_6\Big] \\
&&+\frac{\bar N^3}{M_6^4} T^{zz}_6 \frac{1}{\Box_4+i \varepsilon}
\( T^{yy}_6+2\bar N_y T^{yz}_6+\bar N_y^2 T^{zz}_6\)\,. \nn
\ea
The sign of the kinetic term of $\hat \sigma$ in the final action~\eqref{action3} is therefore irrelevant to this analysis, since this mode is clearly pure gauge and brings no instability.
Furthermore, we notice that $\tilde \chi$ only couples to 6D matter in the bulk, and thus $\pi$ is the only scalar mode to be excited when including an external source on the branes only. Since $\tilde \chi$ is not a true 6D degree of freedom, we may as well integrate it out, leaving only an action for $\pi$ with additional source contributions:
\ba
\label{action4}
S_{\pi}&=& \frac{3 \ms}{2}\int^+ \d^6 x \(\bN \pi \Box_5 \pi-\frac 1\bN (\Ln \pi)^2\) +\frac 32\mf \int_{z=0} \d^5 x \; \pi \Box_5 \pi  \nn \\
&&+ \frac 3 4 \mq \int_{z=y=0} \d^4 x \(\frac{3}{2m_6^2}\frac{\lambda}{\mq}-1\) \pi\Box_4 \pi  \nn\\
&&+\int^+ \d^6x\; \bar N \pi\Big[2T^{\;\mu}_{6\; \mu}-\frac{3}{2} \bar \gamma_{AB}T^{AB}_6-2\frac{\partial_\mu\partial_\nu}{\Box_4}T^{\mu\nu}_6\Big] \nn \\
&&+\int^+ \d^6x\;  \frac{\bar N^3}{M_6^4} \( T^{zz}_6 \frac{1}{\Box_4+i\varepsilon} T^{yy}_6-T^{yz}_6 \frac{1}{\Box_4+i\varepsilon} T^{yz}_6\)\,.
\ea
Now the additional source contributions may appear to have a 4-dimensional pole, but this is in fact not the case. Such terms arise whenever 6D perturbations are decomposed into 4D scalar-vector-tensors, and appear non-local because of the inherent non-locality of the split. In Appendix A we demonstrate that these do not contribute a pole to the single-graviton exchange amplitude.

When considering matter on the 3-brane only,  $T^{AB}_6= \bar N^{-1}\delta(y)\delta_+(z)\, \mathcal{T}^{\mu\nu}_4 \delta_\mu^A\delta_\nu^B$, with $\partial_\mu \mathcal{T}^{\mu\nu}_4=0$, the matter Lagrangian is simply
\ba
\mathcal{L}^{'(4)}_{\rm matter}=\frac 12 \pi \mathcal{T}_4\,.
\ea
This agrees with the flat space-time limit, as well as with the decoupling limit. Meanwhile, the Lagrangian for 5D matter is
\ba
\mathcal{L}^{'(5)}_{\rm matter}=\frac 12 \pi \Big[\mathcal{T}_5^{ab}\eta_{ab}-4\frac{\Box_5}{\Box_4}\mathcal{T}_5^{yy}
\Big]\,,
\ea
where we used $\partial_\mu\partial_\nu \mathcal{T}_5^{\mu\nu}=\partial_y^2 \mathcal{T}_5^{yy}$, which follows from 5D stress-energy conservation.

\section{Discussion}
\label{conclu}

In this paper we have demonstrated the perturbative unitarity and stability of the 6D cascading gravity framework. This puts the claim that this model is a consistent infrared modification of gravity on a firm footing. While we cannot rule out from the previous analysis the existence of an non-perturbative instability, we at least expect a finite range of energies for which this model can be understood as consistent effective field theory. A potentially profound result of this analysis is the confirmation of the result, first obtained in the decoupling limit~\cite{cascade1}, that unitarity requires a minimum tension for the codimension-2 brane localized on an orbifold plane, as soon as we introduce a localized kinetic term for the graviton, {\it i.e.} a 4D induced gravity term. It will be interesting to explore whether a similar bound on the tension applies to higher-codimension branes, and whether there is any fundamental reason for this bound, or whether it is possible to relax this condition by other means (see \cite{gigashif,cascade2}). In any case, our results establish this model as a consistent framework
within which to explore the phenomenology of infrared theories of modified gravity and to confront its predictions against cosmological observations.

\acknowledgments

We would like to thank Gia Dvali, Gregory Gabadadze, Kurt Hinterbichler, Stefan Hofmann, Kazuya Koyama, Oriol Pujolas, Michele Redi, Mark Trodden and Daniel Wesley for useful discussions. A.~J.~T. would like to thank the University of Geneva and LMU, Munich for hospitality whilst this work was being completed. The work of A.~J.~T. at the Perimeter Institute is supported in part by the Government of Canada through NSERC and by the Province of Ontario through MRI. C.~dR is supported by the SNF. The work of J.~K. is supported in part by funds from the University of Pennsylvania and NSERC of Canada.

\section{Appendix A: The decoupling argument}
\label{decoup}

We briefly review the analysis of~\cite{cascade1} using the decoupling limit.
(Note the slight notational differences compared to~\cite{cascade1}. In particular, we have interchanged the roles of $\tilde{h}_{ab}$ and $h_{ab}$ below.)
In the decoupling limit $M_5,M_6\rightarrow \infty$ with $\Lambda_6 = (m_6^4M_5^3)^{1/7}$ fixed, the action~(\ref{6Dcascade}) reduces
to a local field theory on the orbifold brane, describing 5D weak-field gravity and a self-interacting scalar field $\pi$:
\ba
	S_{\rm decoup}^{\rm Jordan} & = & \frac{M_{5}^{3}}{2} \int {\rm d}^{5}x \left[ -\frac{1}{2} h^{ab}(\mathcal{E}h)_{ab} + \pi\eta^{ab}(\mathcal{E}h)_{ab} - \frac{27}{16m_6^2} (\partial_a\pi)^{2} \Box_{5}\pi \right] \nonumber \\
	& & + \int_{y=0} {\rm d}^{4}x \left[ -\frac{M_{4}^{2}}{4} h^{\mu\nu}(\mathcal{E}h)_{\mu\nu} + \frac{1}{2} h^{\mu\nu}T_{\mu\nu} \right]\,,
\label{5Dcov1}
\ea
where $\left(\mathcal{E}h\right)_{ab}  =  -\frac{1}{2} \left(\Box_{5} h_{ab} - \eta_{ab}\Box_{5} h -\partial_a \partial^c h_{cb} - \partial_b\partial^c h_{ac} +\eta_{ab}\partial^c\partial^d h_{cd}+\partial_a\partial_b h\right)$ is the linearized Einstein tensor in 5D, and $(\mathcal{E}h)_{\mu\nu}$ that in 4D. All other interactions in~(\ref{6Dcascade}) are suppressed by powers of $1/M_5$, $1/M_6$ and therefore drop out in the decoupling limit.

With the action written in terms of the physical metric $h_{ab}$, as in~(\ref{5Dcov1}), we notice that $\pi$ is non-minimally coupled to the linearized Ricci scalar, {\it i.e.}, $h_{ab}$ is
the Jordan-frame metric. The kinetic matrix for the metric and $\pi$ can be diagonalized as usual by shifting to the Einstein frame variable, $\tilde{h}_{ab} = h_{ab} -\pi \eta_{ab}$.
Expressed in terms of $\tilde{h}_{ab}$, the decoupling action becomes
\ba
	S_{\rm decoup}^{\rm Einstein} & = & \int {\rm d}^{5}x \left[ -\frac{M_5^3}{4} \tilde{h}^{ab}(\mathcal{E}\tilde{h})_{ab}  -  \frac{3M_5^3}{2} (\partial_a\pi)^{2}\left(1+ \frac{9}{16m_6^2} \Box_{5}\pi\right) \right] \nonumber \\
	& & + \int_{y=0} {\rm d}^{4}x \left[ -\frac{M_{4}^{2}}{4} h^{\mu\nu}(\mathcal{E}h)_{\mu\nu} + \frac{1}{2} \tilde{h}^{\mu\nu}T_{\mu\nu}  + \frac{1}{2}\pi T\right]\,.
\label{5Dcov2}
\ea
As usual, the transformation to Einstein frame generates a conformal coupling of $\pi$ to matter.

In the presence of tension on the 3-brane, $T_{\mu\nu} = -\lambda \eta_{\mu\nu}$, the solution to the equations of motion is
\be
\pi^{(0)} = \frac{\lambda}{3M_5^3}|y|\;; \qquad \tilde{h}_{\mu\nu}^{(0)} = - \frac{\lambda}{3M_5^3}|y|\eta\mn \; ; \qquad \tilde{h}_{y y}^{(0)} = 0\,.
\label{backlamb}
\ee
These combine to give a flat Jordan-frame metric, $h_{\mu\nu}^{(0)} = \tilde{h}_{\mu\nu}^{(0)} + \pi^{(0)}\eta_{\mu\nu} = 0$, consistent with the fact
in 6D that the codimension-2 source generates a deficit angle but leaves the geometry Riemann flat.

To disentangle the different degrees of freedom, let us define
\be
\tilde{h}_{ab}\d x^a \d x^b =\tilde{h}_{\mu\nu} \d x^{\mu} \d x^{\nu}+V\d y^2+2V_{\mu}\d x^{\mu}\d y\,.
\ee
Substituting into the action (\ref{5Dcov2}) gives
\ba
S_{\rm decoup}^{\rm Einstein} & = & \int {\rm d}^{5}x \left[ \frac{M_5^3}{8} \(2V(\partial^{\mu}\partial^{\nu}\tilde h_{\mu\nu}-\Box_4 \tilde h_{(4)}) +4V^{\mu}( \partial_y \partial_{\mu}\tilde h_{(4)}-\partial_y \partial^{\nu}\tilde h_{\mu \nu} ) \right. \right.   \nn \\
&&\left. \left. +\tilde h^{\mu\nu}\Box_5 \tilde h_{\mu\nu}-\tilde h_{(4)}\Box_5 \tilde h_{(4)}+2 \tilde h_{(4)}\partial^{\mu}\partial^{\nu}\tilde h_{\mu\nu}+2(\partial^{\mu}\tilde h_{\mu\nu})^2 -F\mn^2\)\right.\nn\\
&&\left.-  \frac{3M_5^3}{2} (\partial_a\pi)^{2}\left(1+ \frac{9}{16m_6^2} \Box_{5}\pi\right) \right] \nonumber \\
	& & + \int_{y=0} {\rm d}^{4}x \left[ -\frac{M_{4}^{2}}{4} h^{\mu\nu}(\mathcal{E}h)_{\mu\nu} + \frac{1}{2} \tilde{h}^{\mu\nu}T_{\mu\nu}  + \frac{1}{2}\pi T\right]\,,
\ea
with $F\mn=\partial_\mu V_\nu-\partial_\nu V_\mu$.
Concentrating on scalar perturbations only, we may choose a gauge in which the Einstein metric is pure trace: $\tilde h_{\mu\nu}=\frac{1}{4} \tilde h_{(4)} \eta_{\mu\nu}$. However, now it is clear that $V$ and the scalar part of $V_{\mu}$ are Lagrange multipliers which enforce the constraint $\Box_4 \tilde h_{(4)}=0$. Thus in the scalar sector we have $\Box_4 h_{\mu\nu}=\Box_4 \pi \eta_{\mu\nu}$ and $\Box_4 \tilde h_{\mu\nu}=0$. Note that these equations are consistent with the above background solution, at least for $|y|>0$.
The action therefore reduces to the simple form (neglecting the vector part of $V_{\mu}$ which does not couple to 4D matter)
\be
\label{decouplingpi}
S_{\rm decoup}^{\pi} = \int  {\rm d}^{5}x \left[   -  \frac{3M_5^3}{2} (\partial_a\pi)^{2}\left(1+ \frac{9}{16m_6^2} \Box_{5}\pi \right)  \right]  + \int_{y=0} {\rm d}^{4}x \left[ -\frac{3M_{4}^{2}}{4} \pi \Box_4 \pi + \frac{1}{2}\pi T\right]\,.
\ee
Expanding~(\ref{decouplingpi}) to quadratic order in perturbations around the background~(\ref{backlamb}), we find that the $\pi$ cubic term yields a contribution
to the kinetic term of the perturbations that is localized on the 3-brane:
\be
\Delta S_5 = - \int {\rm d}^{5}x \, \frac{9\lambda}{8m_6^2}(\partial_{\mu}\hat{\pi})^2\delta(y)\,,
\ee
where $\hat{\pi}$ now denotes a perturbation around the background profile $\pi^{(0)}$. Combining with the rest of~(\ref{5Dcov2}), the quadratic action is thus given by
\be
\label{decouplingpi2}
S_{\rm decoup}^{\pi} = \int  {\rm d}^{5}x \left[   -  \frac{3M_5^3}{2} (\partial_a \hat \pi)^{2}  \right]  + \int_{y=0} {\rm d}^{4}x \left[ +\frac{3M_{4}^{2}}{4} \(\frac{3\lambda}{2m_6^2M_4^2}-1 \)\hat \pi \Box_4 \hat \pi + \frac{1}{2}\hat \pi \delta T\right]\,.
\ee
This is now manifestly equivalent to the decoupling limit of our final result, reproducing the same minimum bound on the tension.


\section{Appendix B: No extra pole in $TT$ amplitude}
\label{sigma1}

In the final action for $\pi$, given in~(\ref{action4}), we obtained the following term
\be
W\equiv \int^+ \d z \int \d^5 x \frac{\bar N^3}{M_6^4} \( T^{zz}_6 \frac{1}{\Box_4+i\varepsilon} T^{yy}_6-T^{yz}_6 \frac{1}{\Box_4+i\varepsilon} T^{yz}_6\) \,.
\ee
Superficially this looks like a contribution to the $TT$ amplitude which has a 4 dimensional pole and will in general be ghostly. However, since this term already arises in Minkowski space-time it is clear that this cannot be the case. This non-local term arises because the scalar-vector-tensor decomposition is inherently nonlocal.

Absence of ghosts is equivalent to the condition that
\be
{\rm Im} [W] \geq 0\,.
\ee
Since the imaginary part comes entirely from the pole at $\Box_4=0$, it is sufficient to consider the form of the stress-energy on the pole surface, {\it i.e.}, in 4-momentum space with $k^2=0$. Since we are looking at scalar perturbations it then follows that on this surface $\partial_{\mu}T^{\mu y}=\partial_{\mu}T^{\mu z}=0$.
Now conservation of energy implies
\ba
\partial_z (\bar{N} T_6^{zz})+\partial_y (\bar{N} T_6^{yz})+\frac{\partial_y \bar N_y}{\bar N}T_6^{yy}=0\,;\\
\partial_z (\bar N^2 T_6^{zy})+ \partial_y (\bar N^2 T_6^{yy})=0\,.
\ea
We can solve these equations as follows
\be
T_6^{yy}=\bar N^{-2}\partial_z^2 U, \quad T_6^{yz}=-\bar N^{-2}\partial_z\partial_y U, \quad T_6^{zz}=\frac{1}{\bar N}\partial_y (\bar N^{-1} \partial_y U)-\frac{\partial_y \bar N_y}{\bar N^4} \partial_z U.
\ee
Computing the relevant piece, and using the fact that the $\mathds{Z}_2$ symmetry implies $T^{zy}_{z=0}=0$, then $\partial_z U=0$ at $z=0$, and so we find when expressed in momentum space
\ba
&& \int^+ \d z \d y\bar N^3 \( T^{zz}_6(-k) T^{yy}_6(k)-T^{yz}_6(-k) T^{yz}_6(k)\) \nn \\
&&=\int^+ \d z \d y\( -\partial_z \(\bar N^{-1} \partial_y U(-k)\partial_z \partial_y U(k) \) -\frac{\partial_y \bar N_y}{2\bar N^3}\partial_z |\partial_z U(k)|^2\) \nn \\
&&=\(\bar N^{-1} \partial_y U(-k)\partial_z \partial_y U(k) \) _{z=0}+\frac{\partial_y \bar N_y}{2\bar N^3}\left\vert\partial_z U(k)\right\vert^2_{z=0}=0\,,
\ea
where $k$ is the on-shell 4-momentum. In other words, conservation of energy and the orbifold condition ensures that there is no anomalous 4D pole in the propagator.

\section{Appendix C: No pole from $\tilde{\sigma}$ in $TT$ amplitude}
\label{sigma2}

Although the stress-energy on the 4-brane is automatically conserved in 5D and 6D, it is worth stressing that the presence of the singularity at $y=z=0$ imposes an extra constraint for the allowed matter on the orbifold brane. Indeed, the 6D conservation of energy condition applied on the 5D stress-energy implies
\ba
\nabla^{(6)}_A\Big[\frac{1}{\bar N} \mathcal{T}_5^{Az} \delta_+(z) \Big] &=&\Gamma^{(6)\, z}_{\ \ AB}\, \mathcal{T}^{AB}_5 \bar N^{-1} \delta_+(z) \nn\\
&=&\frac{\py \Ny}{\bN^3}\mathcal{T}_5^{yy}  \delta_+(z)=\frac{\lambda}{2\ms \bN^2}\mathcal{T}_5^{yy}  \delta(y)\delta_+(z)=0\,.
\ea
In the background gauge~\eqref{6Dbackground}, the 6D conservation of energy imposes the $yy$-component of the 5D stress-energy to vanish at the singularity $\mathcal{T}_5^{yy}\Big|_{y=z=0}=0$. If this is not so, then this must be compensated by a nonzero contribution to $T^{zA}$. This is a surprising condition, and only really becomes of importance because of the need to regulate the background geometry. Although at first sight it would seem to forbid putting a very small tension on the codimension-1 brane, it is very likely that in practice in the presence of a codimension-1 tension the background will be sufficiently modified that this condition is removed.

If we do not make the assumption that $T^{yy}_5=0$ at $y=z=0$, then on coupling to the metric induced on the orbifold brane, it would appear
that $\hat \sigma$, which decouples from 6D conserved matter, is nevertheless sourced by a term of the form
\be
\int \d^5 x - \frac{\partial_y \bar N_y}{\bar{N}^2} \tilde{\sigma} {\cal T}_5^{yy}=-\int \d^4 x \frac{\lambda}{M_6^4 \bar{N}} \hat{\sigma} {\cal
T}^{yy}_5\,.
\ee
Since $\hat{\sigma}$ enters with the wrong sign kinetic term, such a coupling would seem to indicate a violation of unitarity.
Integrating out $\hat{\sigma}$ this piece generates a term of the form
\be
\int \d^5x\, \delta(y) \frac{\lambda}{2} {\cal T}^{yy}_{5}\frac{1}{\Box_4+i\varepsilon}{\cal T}^{yy}_{5}=\int \d^4 x\, \frac{\lambda}{2} {\cal T}^{yy}_{5}\frac{1}{\Box_4+i\varepsilon}{\cal T}^{yy}_{5}\,.
\label{finaleq}
\ee
However, following the same reasoning as in Appendix B, since the unitarity violation comes entirely from the pole, we may look at the implications of the 5D conservation
equations on the pole surface $k^2=0$. Since we are looking at scalar perturbations it again follows that $\partial_{\mu} {\cal T}_5^{y \mu}=0$
on this surface, and so we have
\be
\partial_y {\cal T}^{yy}_5=0\,.
\ee
For a localized source we may integrate~(\ref{finaleq}) by parts to rewrite it as
\be
{\rm Im}\( \int \d^5x \delta(y) \frac{\lambda}{2} {\cal T}^{yy}_{5}\frac{1}{\Box_4+i\varepsilon}{\cal T}^{yy}_{5}\)=-{\rm Im} \( \int d^5x \epsilon(y)  {\cal T}^{yy}_{5}\frac{1}{\Box_4+i\varepsilon}\partial_y {\cal T}^{yy}_{5} \)=0\,.
\ee
Thus there is no pole contribution to the $TT$ amplitude coming from $\hat{\sigma}$, further confirming that its apparent
wrong sign kinetic term does not indicate the presence of a ghost.


\end{document}